\documentclass{article}

\usepackage{amsmath,amssymb}  
\usepackage{graphicx,color}

\begin{document}

\title{Clustering methods and Bayesian inference for the analysis of 
the  evolution of  immune disorders}
\author{A. Carpio \thanks{Universidad Complutense de Madrid, Spain}, 
A. Sim\'on \thanks{Universidad Complutense de Madrid and Universidad de Granada, Spain}, 
L.F. Villa \thanks{Hospital Universitario Puerta de Hierro de Madrid, Spain}}
\maketitle

{\bf Abstract.}  Choosing appropriate hyperparameters for unsupervised clustering algorithms 
 could be an optimal way for the study of long-standing challenges with data, which
 we tackle while adapting clustering algorithms for immune disorder diagnoses. We 
 compare the potential ability of unsupervised clustering algorithms to detect disease 
 flares and remission periods through analysis of laboratory data from systemic lupus 
 erythematosus (SLE) patients records with different hyperparameter choices. To 
 determine which clustering strategy is the best one we resort to a Bayesian analysis 
 based on the Plackett-Luce model applied to rankings. This analysis quantifies the 
 uncertainty in the choice of clustering methods for a given problem.

\section{Introduction}
\label{sec:intro}

Since the early times of the introduction of mathematical methods for 
medical diagnosis \cite{early}, remarkable advances have been made.
Nowadays, the increasing availability of medical data related to all sorts 
of illnesses is fostering the development of machine learning techniques 
\cite{rana} for medical diagnosis and treatment. While neural networks are 
often used for image based diagnosis \cite{bladder,pneumonia}, supervised 
and unsupervised clustering techniques \cite{oldcluster} are now widely 
employed to investigate the role of genes in sickness 
\cite{genecancer,genecancer2,homology,genearthritis}
and to study the response to therapies \cite{drugcancer,treatmentbone}, 
as well as for assisted clinical diagnosis using information 
from digital devices \cite{accuracy,diagnosisgame}. Developing tools to 
assess the reliability of such automatic procedures and to choose the best 
method for different situations and clinical environments
has become essential \cite{accuracy}.

We consider here the applicability of unsupervised clustering techniques
to identify stages  in time-dependent series of clinical data.
More precisely, we focus on the study of immune disorders, such as SLE,
difficult to diagnose and treat properly because many symptoms are non 
specific and change throughout the evolution of the disease.
SLE is a chronic autoimmune disease in which the immune system attacks 
healthy tissues by mistake \cite{lupus}.
This attack causes inflammation and, in some cases, permanent tissue damage.  
Parts of the body commonly affected include skin, joints, heart, lungs, kidneys, 
red bone marrow (blood cell formation) and brain. Symptoms of lupus vary 
between people and may be mild to severe, affect 
one area of the body or many, come and go, and change over time. They include 
painful and swollen joints, fever, fatigue, chest pain when breathing deeply, hair 
loss, mouth ulcers, swollen glands, a red rash (typically on the face)
and cardiac, renal o neural symptoms.

The cause of SLE is unknown. It is thought to involve genetics as well as environmental 
factors. Female sex hormones, race, age between 15 and 45 and family history appear 
to be risk factors for SLE development. Aside, emotional or physical stress, sunlight  
exposure, viral infections, certain drugs, pregnancy, giving birth, smoking or vitamin 
D deficiency can trigger disease flares. There is currently no cure for lupus. 
Treatments include nonsteroidal anti-inflammatory drugs, corticosteroids, immunosuppressants, hydroxychloroquine, and immunomodulatory monoclonal antibodies.
However, dealing with a chronic disease makes one concerned about long-term adverse 
effects. Several treatments for lupus have attracted great attention recently due to their 
applicability  to covid-19 patients. In fact, serious covid-19 cases develop similar hyperimmune 
responses  \cite{covid} which damage the patient's tissues and may cause death.

Lupus patients go through periods of illness, called flares, and periods of wellness, 
called  remission. The symptoms during flares vary. It is essential to be able to 
distinguish early when the patient is transitioning from remission to flares, as well as 
what factors are causing it. A great deal of information is contained in laboratory tests. 
Our purpose is to develop mathematical methods to process it automatically from time 
series of such tests. We show here that it could be possible to automatically detect 
transition days  by applying clustering techniques. Different clustering strategies may produce 
variable results on the same datasets. Therefore, it is important to be able to assess 
which algorithms perform better in given tests problems.  We show that a Bayesian 
analysis of rankings of the performance of clustering algorithms on medical datasets 
provides information on the probability of each clustering algorithm being the best.

The rest of the paper is organized as follows. Section \ref{sec:data} describes
the datasets under study.  Sections \ref{sec:kmeans}, \ref{sec:hierarchical}
and \ref{sec:dbscan} apply K-means, Hierarchical clustering and Density
based spatial clustering  to the selected datasets. Section \ref{sec:bayesian}
explains how to construct performance rankings to estimate the probability of
a particular clustering procedure and hyperparameter choice to be the most 
adequate one to identify automatically transition from remission to flares and 
other stages in the patient evolution from time series of clinical data. We 
summarize our conclusions in Section \ref{sec:conclusion}. A final Appendix 
details the clinical variables under consideration.


\section{Clustering clinical data}
\label{sec:data}

Laboratory tests are often used to diagnose lupus, since the illness involves an 
immune response by antibodies against the patients's own body.  
In addition, laboratory tests play a key role when detecting the
transition to flares and identifying the kind of health disorder that is building up 
and needs to be treated.  Tables \ref{TableAp1}-\ref{TableAp3} in the Appendix 
list some variables usually monitored. We will work here with $28$ patient records, 
choosing one dataset to illustrate the outcome of the clustering procedures and all 
datasets for the Bayesian study of the probability of a specific method being the 
best to identify illness stages in these patient's records.

\begin{figure}[!h] 
	\centering
	\includegraphics[width=13cm]{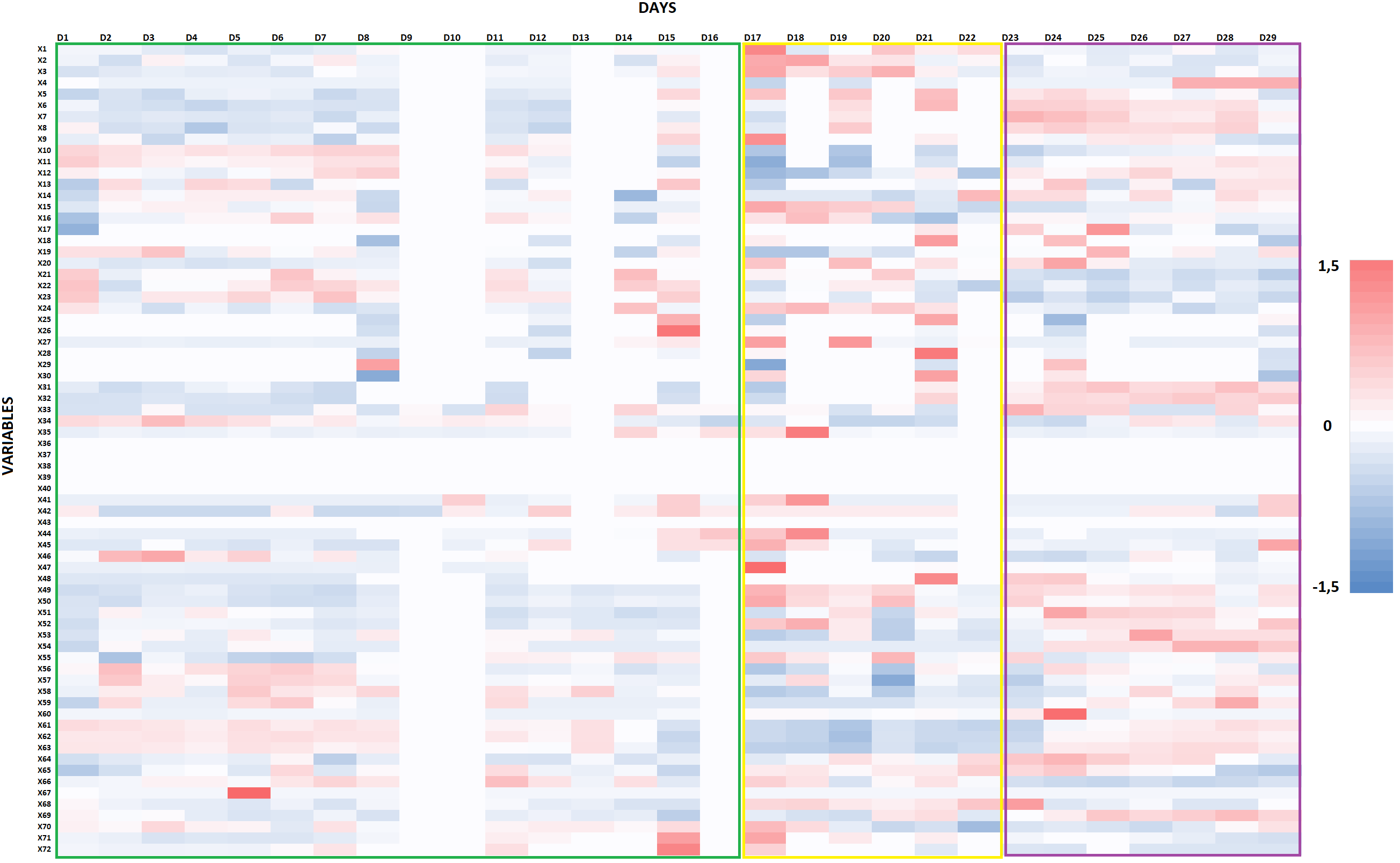}
	\caption{Heatmap of the reference time series of clinical data. 
	} 
	\label{Imagen:Figura12}
\end{figure}

\begin{figure}[!h] 
	\centering
	\includegraphics[width=12cm]{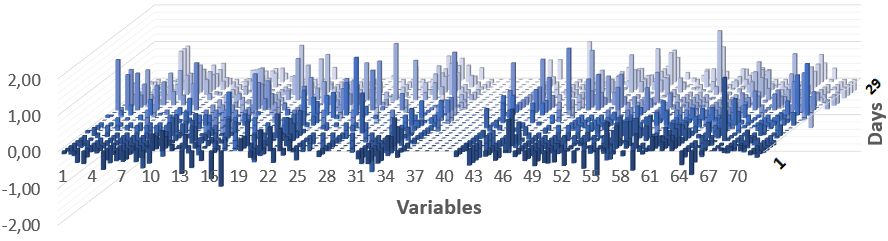} 
	\caption{Bar histogram of the reference clinical data. 
	} 
	\label{Imagen:Figura13}
\end{figure}

All datasets are normalized subtracting the mean from each variable and dividing
by three times the variance. After that, we obtain normalized time series of clinical  
data as represented in Figure \ref{Imagen:Figura12}.
A difficulty in dealing with time series of clinical data is that some measurements
are usually missing. White boxes mark missing data. We will eliminate from our
study all variables with more than $50\%$ measurements missing. For the remaining  
variables, we fill empty boxes with the average value of the variable over the remaining 
days. Variables labeled as $17$, $18$, $28$, $29$, $30$, $47$, $48$ in Tables 
\ref{TableAp1}-\ref{TableAp3} are suppressed.
At the end, our data set is formed by $29$ columns (days) and $65$ rows (variables).
From this set, we may also eliminate six more variables with essentially zero values.
Time measurements are not consecutive (they are not recorded in consecutive days) 
but expand over months. The label attached to the days only indicates time ordering.
Visually, we can identify three groups of days. Group 1 is formed by days $1$ to $16$, 
group 2 is formed by days $17$ to $22$, and group 3 is formed by days $23$ to $29$.
The complementary representation in Figure \ref{Imagen:Figura13} shows how certain 
variables get out out control as time advances, and later start coming back to normal, 
responding to treatment. These observations provide the motivation to look for clusters 
in clinical data.

Automatic clustering techniques may produce clusters even when the original data 
contain no significative clusters. Hopkins criterion \cite{Hopkins} allows us to establish 
whether the dataset contains relevant clusters. 
Given a set $D$, the Hopkins  statistics is obtained as follows:
\begin{enumerate}
\item We extract a uniform sample  $(p_1,\dots, p_n)$ from $D$ formed by $n$ points.
\item For each point $p_i \in D$, we find the closest neighbor $p_j$ and denote the distance  
$x_i = dist(p_i, p_j)$.
\item We generate a random sample $(q_1,\dots, q_n)$ with $n$ points, which we call 
$D_{random}$ from a uniform distribution keeping the same variance as the original set $D$.
\item For each $q_i \in D_{random}$, we find its closest neighbor $q_j$ in $D$ and denote 
$y_i = dist(q_i, q_j).$
\item We calculate the Hopkins  statistics   as
\begin{eqnarray*}
H = \frac{\sum_{i=1}^{n} y_i}{\sum_{i=1}^{n} x_i + \sum_{i=1}^{n} y_i}.
\end{eqnarray*}
\end{enumerate}
If $D$ was uniformly distributed, then the sums $\sum_{i=1}^{n} y_i$ and $\sum_{i=1}^{n} x_i$ 
would be similar, and $H$ would be close to $0.5$. However, if there are clusters present in 
$D$, the distances to the artificial points $(\sum_{i=1}^{n} y_i)$ would be larger than4 those 
to the true points $(\sum_{i=1}^{n} x_i)$ and $H$ would be larger than $ 0.5$. The larger is 
$H$, the more likely the presence of clusters in the data is.

For our dataset, $H = 0.655$, so that it makes sense applying clustering techniques.
An additional issue when dealing with medical data, is whether such clusters have a 
medical meaning, which will be the case here.


\section{K-means clustering}
\label{sec:kmeans}

K-Means \cite{Kmeans} is one of the most widely used unsupervised clustering
algorithms. The idea is to group observations in clusters so that the total intra-cluster 
variation is minimized.

\subsection{Algorithm description}
\label{sec:kmeans_algorithm}

Given an observation $\mathbf x_i=(x_{i,1}, \ldots, x_{i,M})$ in a $M$ dimensional space, 
a cluster $C_j$ of points in the same space and the cluster centroid $\boldsymbol \mu_j$, 
we define the intra-cluster total variation as:
\begin{eqnarray*}
\sum_{j=1}^{K} W(C_j) = \sum_{j=1}^{K} \sum_{\mathbf x_i \in C_j} d(\mathbf x_i,
\boldsymbol \mu_j),
\end{eqnarray*}
where $d$ represents the euclidean distance 
\begin{eqnarray*}
d(\mathbf x_i,\boldsymbol \mu_j) = \sqrt{\sum_{\ell=1}^M (x_{i,\ell} - \mu_{j,\ell})^2}.
\end{eqnarray*}
The centroids of each cluster $C_j$ with $|C_j|$ observations are defined as 
the averages of the observations in the cluster, that is, 
$\boldsymbol \mu_j = \sum_{\mathbf x_i \in C_j} \mathbf x_i/|C_j|.$
The intra-cluster total variation can be interpreted as a measure of the cluster
compactness. Each term $W(C_j)$ is the intra-cluster variation for a single
cluster:
\begin{eqnarray*}
W(C_j) = \sum_{\mathbf x_i \in C_j} d(\mathbf x_i,\boldsymbol \mu_j),
\end{eqnarray*} 
where $\mathbf x_i$ are the points belonging to the cluster $C_j$. In our case, each 
observation is formed by measurements of $M$ clinical variables a specific day.

Once the number $k$ of clusters to be formed is specified, the K-Means algorithm proceeds 
in the following steps:
\begin{enumerate}
\item We initialize the centroids $\boldsymbol \mu_j$ generating generate $k$ random  
points in the  $M$ dimensional space.
\item Each observation $\mathbf x_i$ is assigned to the closest centroid according to 
the euclidean distance.
\item For each cluster, we update the centroid as the average of the cluster observations.
\item We minimize the total intra-cluster variation iteratively. To do so, we iterate the 
previous steps until the clusters do not change or the maximum number of iterations 
is surpassed.
\end{enumerate}
The main drawback of this algorithm is the need of knowing the number of clusters 
beforehand. We describe some strategies to estimate it next.

\subsection{Selection of the number of clusters}
\label{sec:kmeans_tuning}

Two methods can be used to estimate the number of clusters in K-Means: 
the Elbow method  and the Silhouette method. 
While the Elbow method favors cluster compactness, the Silhoutte analysis 
opts for cluster separation when selecting $k$.

\begin{figure}[!h]  
	\centering
	\hskip 5cm (a) \hskip 6cm (b) \\
	 \includegraphics[width=6cm]{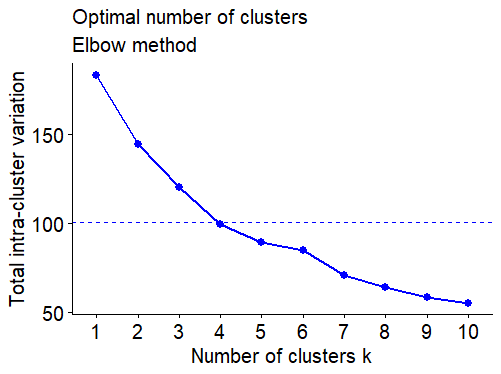} 
	 \includegraphics[width=6cm]{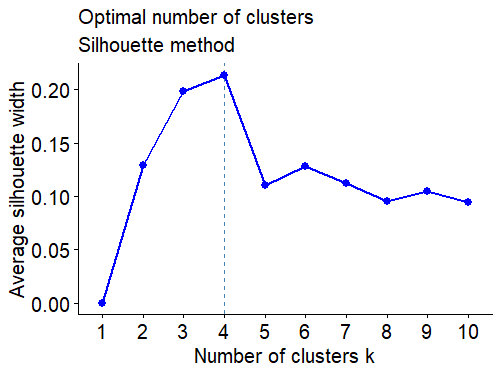} \\
	\caption{(a) Elbow Method: Change in the total intra-cluster variation as a
	function of the number of clusters $k$. (b) Silhouette method: Silhouette 
	coefficient as a function of the number of clusters $k$.
	} 
	\label{Imagen:Figura6}
\end{figure}

The Elbow method is based on running K-Means for different choices of the number 
of clusters. Each run stores the total intra-cluster variation. The number of clusters 
$N$ is selected in such a way that the total intra-cluster variation does not diminishes 
noticeably for $k+1$. This can visualized graphically, plotting the total intra-cluster 
variation as a function of $k$. When applied to our reference clinical dataset, we find 
Figure \ref{Imagen:Figura6}(a), which suggests  $k=3,4,5$ as reasonable values.

The Silhouette analysis \cite{Silhouette} measures cluster quality. This method determines 
the quality of a cluster estimating how well the point fits in the cluster. This is done calculating 
the mean distance from each point to the other clusters, the so-called Silhouette coefficient.
Choosing the number of clusters maximizing the Silhouette coefficient we guarantee a sharp 
separation between clusters. Figure \ref{Imagen:Figura6}(b) represents the Silhouette coefficient as a function of $k$ for our reference dataset. The value maximizing the coefficient is $k=4$,  though $k=3$ is only slightly worse.

\subsection{K-Means applied to a time series of clinical data}
\label{sec:kmeans_application}

\begin{figure}[!h] 
	\centering
	\includegraphics[width=8cm]{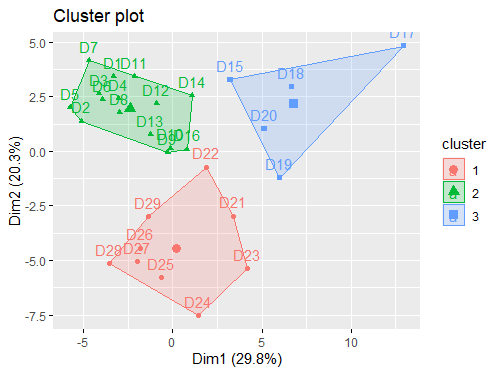} 
	\caption{Clusters obtained with K-Means when $k=3$.} 
	\label{Imagen:Figura8}
\end{figure}

\begin{figure}[!h] 
	\centering
	\includegraphics[width=8cm]{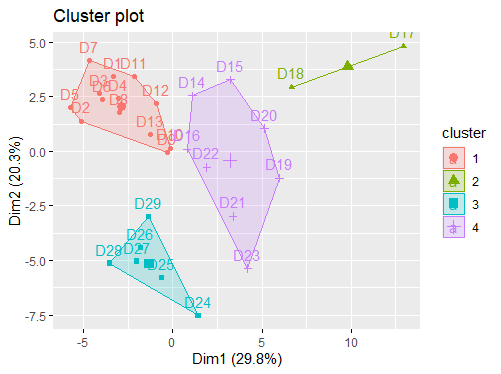} 
	\caption{Clusters obtained with K-Means when $k=4$.  } 
	\label{Imagen:Figura9}
\end{figure}

The results obtained running K-Means for $k=3$ and $k=4$ clusters are represented in 
Figures \ref{Imagen:Figura8} and \ref{Imagen:Figura9}.

As Figure \ref{Imagen:Figura8} shows, the algorithm is able to identify groups
of days, which are almost consecutive.
The first (red) group includes days $21$ to $29$.  
The second (green) group includes days $1$ to $16$, skipping day $15$.  
Finally, the third (blue) group includes days $17$ to $20$, and also day $15$. 
Notice that day $15$ is an anomaly for K-Means. 

Let us see the behavior for $k=4$. Again, Figure \ref{Imagen:Figura9} shows, 
that the algorithm is able to identify groups of days.
The first (red) group includes days $1$ to $13$.  
The second (green) group includes days $17$ and $18$.
The third (blue) group includes days  $24$ to $29$.  
Finally, the fourth (magenta) group includes days $14$ to $16$, plus days $19$-$23$.
This time days $17$-$18$ are an anomaly for K-Means. 

As a conclusion, days $15$, $17$ and $18$ are difficult to explain for this algorithm.
This may mean that they are days at which the patients' condition changes.

\section{Hierarchical clustering}
\label{sec:hierarchical}  

Hierarchical clustering  \cite{agnes} is a popular strategy for unsupervised learning. 
We will use here the agglomerative version of the algorithm, which works bottom
up. Each element is initially considered a cluster itself. A cluster formed by a single 
element is a leaf. At each step, the two most similar clusters merge to form a bigger 
one, called node. This process is repeated progressively until all the points are 
combined in a big cluster, the tree.

\subsection{Algorithm description}
\label{sec:hierarchical_algorithm}

This algorithm requires a similarity measure between elements and a strategy
to merge clusters. Different distances can be selected as similarity measures. Here 
we use the euclidean distance. The elements to be compared are the values of  the 
$M$ variables for different days. Considering them points in a $M$ 
dimensional space, we can compute the euclidean distance between them. Regarding  
the clustering strategy, we calculate the distances between all the elements and use 
the 'complete' approach.
Each node's height in the tree  represents the distance between the two subnodes  
merged at that node.  All leaves below any node whose height is less than a threshold 
$C$ are grouped into a cluster (a singleton if the node itself is a leaf).
This process is represented in a dendrogram, a graph representing how the
clusters merge until they form the tree that contains them all. As we move upwards 
the most similar clusters combine in branches that merge later at a higher height, 
known as the cophenetic distance between the clusters.
The higher that height, the more different the clusters are. 

Once a hierarchical tree is built, we have to check whether it is representative
of our set of data, that is, whether the heights represent the original distances
with reasonable accuracy. To do so, we calculate the correlation between the
cophenetic distance and the original distance used to check similarity between
objects, the euclidean distance in our case. If the correlation coefficients
displays a large coefficient in a linear relation, say, larger than $0.75$, the tree
is considered a good representation of the dataset. Selecting a particular
height to cut the tree, we obtain different numbers of clusters.

\subsection{Hierarchical clustering applied to a time series of clinical data}
\label{sec:hierarchical_application}

\begin{figure}[!h] 
	\centering
	\includegraphics[width=8cm]{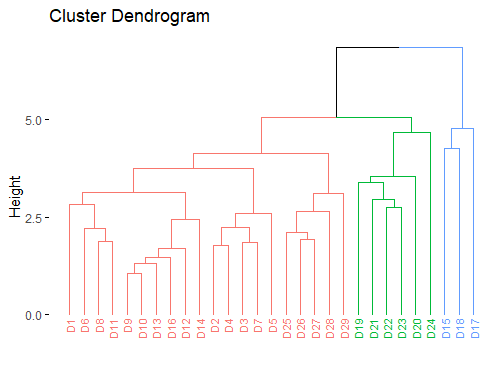} 
	\caption{Dendrogram representing $3$ clusters in the time
	sequence of clinical variables} 
	\label{Imagen:Figura15}
\end{figure}

\begin{figure}[!h] 
	\centering
	\includegraphics[width=8cm]{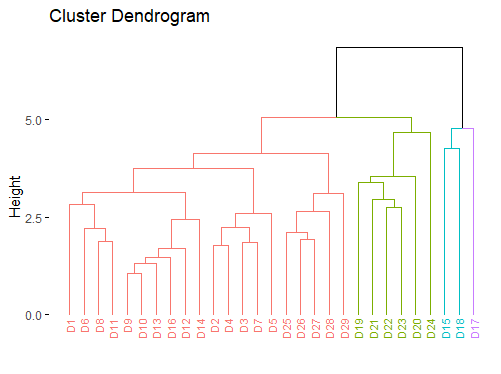} 
	\caption{Dendrogram representing $4$ clusters in the time
	sequence of clinical variables} 
	\label{Imagen:Figura16}
\end{figure}

In this section we  use the hierarchical clustering to select the onset of severe 
illness periods in time series of measurement of the clinical variables of lupus 
patients. 
The dendrograms in Figure \ref{Imagen:Figura15} and \ref{Imagen:Figura16}
show the outcome of applying agglomerative hierarchical clustering based on 
the euclidean distance cut at a different height. The resulting tree is a good 
representant  of the data set, since the correlation
between the cophenetic distance and the euclidean distance is $0.7767 > 0.75$.
We select the hyperparameter, that is, the height, is such a way that we obtain the 
number of clusters we considered with K-means.
In the first case, choosing a threshold height to have $3$ clusters,  days
$15$, $17$, $18$ are singled out. 
In the second one, with $4$ clusters, day $17$ is marked as possible onset.
Days $15$, $17$, $18$ are automatically identified as days at which the patients 
condition may change significantly again.

\subsection{Grouping clinical variables}
\label{sec:kmeans_variables}

Both K-Means and hierarchical clustering select the same days to mark a 
strong alteration in the status of a patient. We can analyze the clinical variables 
by a combination of both.

\begin{figure}[!h] 
	\centering
	\includegraphics[width=13cm]{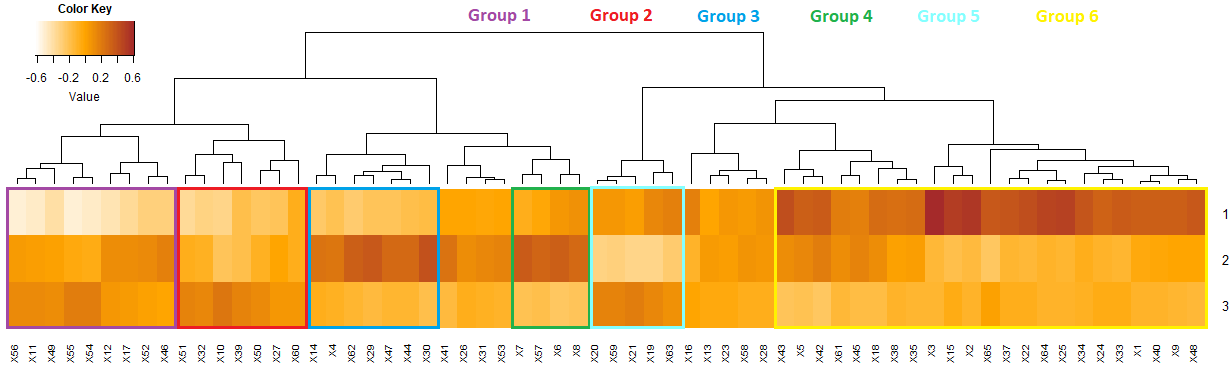} 
	\caption{Heatmap of the matrix relating the K-Means centroids for the $k=3$ clusters 
	of days and the clinical variables. We have superimposed a dendrogram classifying
	the variables in groups according to their influence in the cluster centroids.}
	\label{Imagen:Figura11}
\end{figure}

K-Means algorithm provides a matrix describing how each variable influences the 
final centroids. We represent this matrix by means of a heatmap for $k=3$ clusters,
see Figure \ref{Imagen:Figura11}. Then, we superimpose a dendrogram
grouping the variables by similarity. In this way, we distinguish 6 groups of variables 
according to their interactions through the clusters.
Considering high values the largest positive ones, intermediate values those about 
$0$ and small values the largest negative ones, we observe:
\begin{itemize}
\item The low values of group 1 identify the first cluster, while intermediate values
are typical of the second and third clusters.
\item The low values of group 2 identify the first cluster, while intermediate values
are typical of the second cluster and large values of the third cluster.
\item The large values of group 3 identify the second cluster, while small values
are typical of the first and second clusters.
\item The large values of group 4 identify the second cluster, while small values
are typical of the third cluster and intermediate values of the first cluster.
\item The low values of group 5 identify the second cluster, while intermediate-large 
values are typical of the first  and  third clusters.
\item The large values of group 6 identify the first cluster, while intermediate-small 
values are typical of the second and third clusters.
\end{itemize}
A few variables remain almost constant through the clusters and we have left them unassigned. This shows that they have little effect on the overall clustering results and
we may as well suppress them.
These relations are illustrated in Table \ref{Table1}. 
Variables for which the heat map reports large positive values within the cluster affect 
strongly  the centroid: its coordinate in the direction of that variable is large.
Considering variables for which the heat map reports small values (that is, large negative values) within the cluster, the centroid varies strongly in the opposite direction.
In practice, this information would allow to track automatically which variables
play relevant roles in each cluster of days and motivate the change.

\begin{table}[!h]
\centering
\begin{tabular}{|c|c|c|c|}
\hline
                 & \textbf{Large} & \textbf{Intermediate} & \textbf{Low} \\ \hline
\textbf{Group 1} & 2-3          & 2-3             & 1             \\ \hline
\textbf{Group 2} & 3             & 2                & 1             \\ \hline
\textbf{Group 3} & 2             & 1-3             & 1-3           \\ \hline
\textbf{Group 4} & 2             & 1                & 3             \\ \hline
\textbf{Group 5} & 1-3          & 1-3             & 2             \\ \hline
\textbf{Group 6} & 1             & 2-3             & 2-3           \\ \hline
\end{tabular}
\label{Table1}
\caption{Relation between day clusters and clinical variable blocks.}
\end{table}


\section{Density-based spatial clustering}
\label{sec:dbscan}
\begin{figure}[!h] 
	\centering
	\includegraphics[width=5cm]{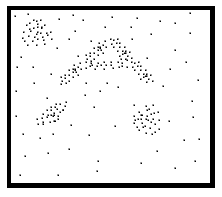} 
	\caption{Density based cluster identification in the presence of noise.
	} 
	\label{Imagen:Figura26}
\end{figure}

The Density-Based Spatial Clustering of Applications with Noise (DBSCAN) algorithm usually allows us to detect clusters of any structure even in the presence of noise and outliers. This technique is based on the spatial density of points, following human intuition to identify clusters. For instance, in Figure \ref{Imagen:Figura26}, we visually spot four clusters in spite of the presence of noise due to the point density variations.

The method looks for high density regions and assigns clusters to them, while points in less dense regions are left outside and become anomalies. This is the main advantage of this algorithm: being able to detect outliers. Another advantage is that methods such as K-Means and Hierarchical Clustering are designed to find spherical or convex clusters, that is, they work well when we must find well separated and compact clusters, which is not always the case. Moreover, K-Means also needs to classify all the points in one cluster forcing the introduction of weird criteria to detect outliers.

\subsection{Algorithm description}
\label{sec:density_algorithm}

\begin{figure}[!h] 
	\centering
	\includegraphics[scale=0.8]{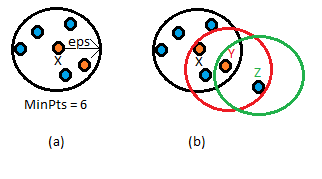} 
	\caption{Cluster detection in the presence of noise by density criteria.
	} 
	\label{Imagen:Figura27}
\end{figure}

The DBSCAN algorithm is governed by two hyperparameters:
\begin{itemize}
\item $\varepsilon$: Smallest distance for two points to be considered neighbors.
\item $MinPts$: Minimum number of points required to form a cluster.
\end{itemize}
According to them, we distinguish three types of points:
\begin{itemize}
\item Core Point: Any point with a number of neighbors greater than or equal
to a fixed minimun value $MinPts$ (including itself).
\item Border Point: Any neighbor of a core point with a number of
neighbors smaller than $MinPts$.
\item Outlier: Any point which is not a core neither a border point.
\end{itemize}
Figure \ref{Imagen:Figura27} illustrates the three types of points for a given
$\varepsilon$ and  $MinPts = 6$. Point $\mathbf{x}$ is a core point, it has at least
$5$ neighbors at a distance smaller than $\epsilon$, a total of $6$ points
counting $\mathbf{x}$. On the other hand,  $\mathbf{y}$ is a border point, since
the number of neighbors is smaller than $6$, but it is a neighbor of the
core point $\mathbf{x}$. Finally, $\mathbf{z}$ is an outlier. Although it is a neighbor  of $\mathbf{y}$, it is not a neighbor of any core point and it has less than $6$  neighbors.

Before describing the algorithm in detail, we need to distinguish three concepts:
\begin{itemize}
\item Directly reachable points: A point $A$ is directly reachable from $B$ when
it is a neighbor of $B$ and $B$ is a core point.
\item Reachable points: A point $A$ is  reachable from $B$ when we can find
a set of core points from $B$ to $A$.
\item Connected point: Two points $A$ and $B$ are connected when there is a core 
point $C$ such that $A$ and $B$ are reachable from $C$.
\end{itemize}
A density based cluster is a group of connected points. The DBSCAN algorithm
works as follows:
\begin{enumerate}
\item For each point $\mathbf x_i$, we calculate the distance between $\mathbf x_i$ and the remaining points. Then, we find all the neighbors within the radius $\varepsilon$ and mark as core points those with a number of neighbors greater or equal to \textit{MinPts}.
\item For each core point not yet assigned to a cluster, we create a new cluster.
Next, we search for all the points connected to that core point and assign them to
that cluster. 
\item We repeat those steps over the remaining set of points.
\item The points not assigned to any cluster after this process are considered outliers.
\end{enumerate}

\subsection{Hyperparameter tuning}
\label{sec:density_tuning}

Determining adequate values for $\varepsilon$ and $MinPts$ is a difficult task, strongly conditioned by the structure of the dataset we are working with. In general, there is no automatic procedure to do so. The main problems caused by a poor choice are:
\begin{itemize}
\item If the value of $\varepsilon$ is too small, there will not be enough points to
form clusters and most points risk being classified as outliers. On the other hand,
if $\varepsilon$ is too large, most points will be classified in clusters and we will
not be able to identify the outliers. 
\item  If $MinPts$ is too large, too many points are required to form a cluster. Dense regions may be classified as outliers. When it is too small, low density regions would appear as clusters and outliers would remain undetected.
\end{itemize}
These values have to be carefully tuned to detect meaningful outliers. Our choice
will be to find optimal values of $\varepsilon$ given $MinPts.$ We start calculating
the means of the distances to the $k$ closest points and represent them in increasing order. Turning points will mark thresholds for sharp changes and will provide candidate values for 
$\varepsilon$. 

\subsection{DBSCAN applied to a time series of clinical data}
\label{sec:density_application}

In this section we explain how to use the clustering algorithm DBSCAN 
\cite{dbscan} to select the onset of severe illness periods in time series of 
measurement of the clinical variables of lupus patients. 

We select $\varepsilon$ fixing the hyperparamenter $MinPts=3$, minimum
number of days in clusters usually observed in our previous studies.
The resulting graph would be
 \begin{figure}[!h] 
 	\centering
 	\includegraphics[width=6cm]{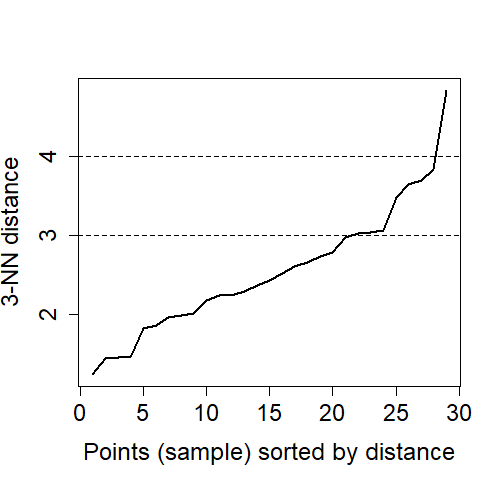} 
 	\caption{Graph of ascending $3$-distances for the reference dataset of medical variables.
	} 
 	\label{Imagen:Figura29}
 \end{figure} 
We appreciate two turning points, at $\varepsilon=3$ and $\varepsilon=4$.  However,
$\varepsilon=4$ is too large and all the points form a cluster. 
Running the algorithm with $\varepsilon=3$ and $\varepsilon=3.5$ we find the following.

\begin{figure}[!h] 
 	\centering
	\hskip 5cm (a) \hskip 6cm (b) \\ \hskip -2mm
	\includegraphics[width=6cm]{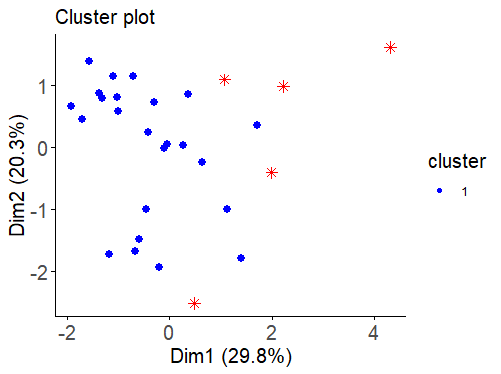} \hskip 2mm
	\includegraphics[width=6cm]{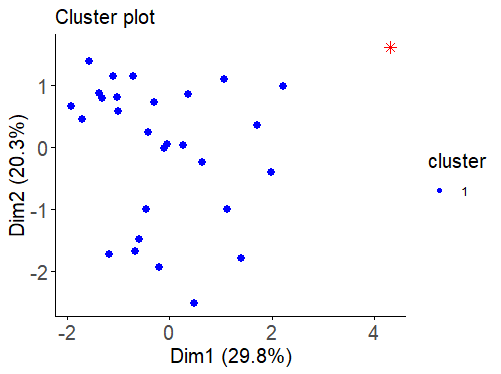} \\
 	\caption{(a) Clusters and outliers with DBSCAN and $\varepsilon=3$.
	(b) Clusters and outliers with DBSCAN and  $\varepsilon=3.5$. 
	} 
 	\label{Imagen:Figura30}
 \end{figure}

For $MinPts=3$ and $\varepsilon = 3$, there are five outliers. The first one (day $15$) marks  the correct onset of the sickness period, see Figure \ref{Imagen:Figura30}(a).
On the other hand, when $MinPts=3$ and $\varepsilon = 3.5$, we find a single outlier (day $17$), two days later, see Figure \ref{Imagen:Figura30}(b). This technique confirms what  we have previously observed: the clinical variables are strongly altered between the days $15$ and $17$.

\section{Bayesian inference for clustering performance}
\label{sec:bayesian}

For the same set of data, results vary using different clustering procedures. Thus, comparing the performance of different clustering strategies to extract useful information from clinical datasets is an important task. In the previous sections, we have illustrated the use of clustering strategies on a dataset to obtain information on the phases of the illness of a patient, in particular, to characterize the onset of deterioration and recovery periods. Next, we describe how to infer in an automatic way which methods are more adequate for particular collections of clinical records by combining the construction of rankings on smaller collections of selected datasets and  Bayesian  analysis. 

\subsection{Bayesian inference for ranking analysis}
\label{sec:bayesian_algorithm}

The idea is to run the clustering methods to be compared on a number of datasets for which the diagnosis is known, and then rank them according to their performance. Finally, we analyze the results using a Plakett-Luce (PL) Bayesian model  \cite{Ceberio,permallow}. This model is well adapted to this type of problems and differs from other approaches for continuous problems \cite{sergei}.

An advantage of PL models is that normalized parameters represent directly the
marginal probability of an algorithm being placed in first position. Another advantage is that is relies on a finite number of parameters, as many as algorithms we compare. It also fulfill's Luce's axiom: 'The probability of an algorithm $A$ of being placed before algorithm $B$ is the same independently of the remaining algorithms'.

The PL model combines three ingredients:
\begin{enumerate}
\item It selects randomly the algorithm to be positioned.
\item Each algorithm has a weight $w_i$, and the probability to select an algorithm at each stage is the ratio between its weight and the sum of the weights of the remaining algorithms.
\item Representing by  $\sigma = (\sigma_1, \dots , \sigma_n)$ a ranking of size $n$, where $\sigma_i = j$ implies that the $j$-th algorithm is locates at the $i$-th position and by $w = (w_1, \dots, w_n)$ the vector of weights, the PL probability to select that ranking is:
\begin{eqnarray*}
 P_{PL}(\sigma) = \displaystyle\prod_{i=1}^n \frac{w_{\sigma_i}}{\sum_{j=i}^n w_{\sigma_j}}. 
\end{eqnarray*}
\end{enumerate}
By simplicity, we assume that the sum of weights is $1$.

\begin{figure}
\centering
\includegraphics[width=12cm]{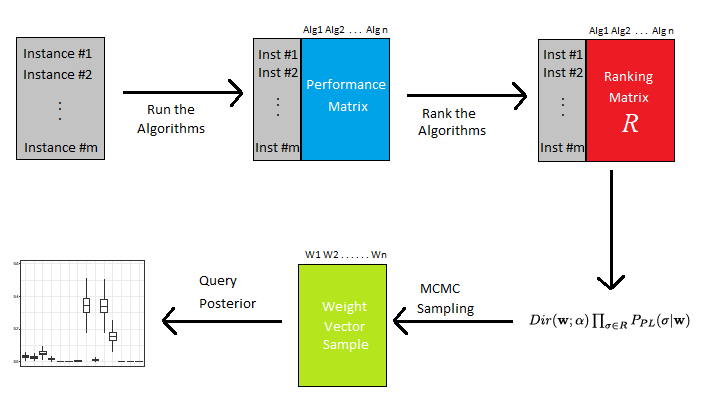} 
\caption{Scheme for the Bayesian analysis of clustering ranking performance.
} 
\label{Imagen:Figura1}
\end{figure}

More precisely,  the process, schematized in Figure \ref{Imagen:Figura1}, is the
following:
\begin{enumerate}
\item Run the clustering algorithms over $m$ datasets. Choosing a way to evaluate the performance of the algorithms, we will have a matrix $\mathbf M$ in which $M(i,j)$ is the performance of algorithm $j$ on dataset $i$.
\item Assign a position in the ranking to each algorithm: the greatest the performance,  the higher the situation in the ranking. In this way, we obtain a matrix $\mathbf  R$, where $\mathbf R(i,j)$ is the location in the ranking of algorithm $j$ applied to dataset $i$.
\item A given matrix $\mathbf R$ gives partial information of the likelihood of different clustering algorithms to perform better than the rest. We quantify the uncertainty in such conclusions using Bayes relation for conditional probabilities
\begin{eqnarray*}
P(\mathbf w| \mathbf  R) \propto P(\mathbf w) \cdot P(\mathbf R|\mathbf  w),
\end{eqnarray*}
where $\mathbf  w=(w_1,...,w_n)$ denote the PL parameters. Since we are assuming $\sum_{i=1}^n w_i=1$, they represent the probability of having a given algorithm in the first position. We set
\begin{eqnarray*}
P(\mathbf  R|\mathbf  w) = \Pi_{\sigma \in \mathbf  R} P_{PL}(\sigma,\mathbf  w), \\
P(\mathbf w) = {\rm Dir}(\mathbf w, \boldsymbol \alpha),
\end{eqnarray*}
where the Dirichlet distribution models uncertainty in the weights. 
\item Since the posterior distribution does not have a closed form, we sample it using Markov Chain Monte Carlo (MCMC) methods \cite{mcmc}. From the set of samples, we visualize uncertainty in the probability of a clustering strategy being the best by means of histograms or expected values.
\end{enumerate}

The Dirichlet distribution Dir($\boldsymbol \alpha$) is a family of multivariate distributions parametrized by a vector $\boldsymbol \alpha$  of real positive numbers. It generalizes the Beta function. The Dirichlet distribution of order $K \geq 2$ with positive parameters $\alpha_1,\dots, \alpha_K$ has a density function 
\begin{equation*}
f(x_1,\dots, x_K;\alpha_1, \dots , \alpha_K) = \frac{1}{B(\boldsymbol \alpha)}
\prod_{i=1}^{K}x_i^{\alpha_i - 1}
\end{equation*}
where $\sum_{i=1}^{K}x_i = 1$ and $x_i \geq 0$ for $i \in [1,K]$. The normalization constant is a multivariate Beta function expressed in terms of Gamma functions as:
\begin{equation*}
B(\boldsymbol \alpha)  = \frac{\prod_{i=1}^{K}\Gamma(\alpha_i)}{\Gamma(\sum_{i=1}^{K} \alpha_i)} \hspace{1cm} \boldsymbol \alpha = (\alpha_1,\dots,\alpha_K).
\end{equation*}
Since we lack information on the perfomance of the algorithms for general datasets, we use a uniform distribution for the  hyperparameter $\boldsymbol \alpha$, that is,
 $\alpha_i = \alpha = 1, i=1,\dots,n$. 

\subsection{Plackett Luce method applied to time series of clinical data}
\label{sec:bayesian_applied}

We apply our clustering techniques to identify groups of days reflecting different
stages of the evolution of the patient. 
We consider the clinical records of $28$ patients which have been previously 
diagnosed. 
In many cases there are sickness periods which require special medical care,
starting a known specific day. This results in a time series of measurements
of clinical variables, which display a different behavior before and after it.
A way to quantify the performance of the
different algorithms of the datasets if to check how well they predict the onset
of that flares period. Therefore, we need to define an automatic criterion to
identify it based on the different clustering strategies:
\begin{itemize}
\item For the DBSCAN algorithm we select the first day not assigned to any cluster, 
that is, the first outlier.
\item For K-Means, we choose the first day which is unsequentially classified, that is, 
which has no neighbors in the same cluster. When such a day is not found, we
choose the first day of the smallest cluster.
\item For hierarchical clustering, we choose the smallest of the last isolated points 
to merge with existing clusters. If not found, we choose the first day within the 
smallest cluster.
\end{itemize}
Performance is quantified by means of the difference between the day estimated from 
the clustering analysis and the known true transition day.

Let us revise the clustering analysis performed on our test dataset and apply these criteria. The onset of an unstability period for the patient is defined by Day $15$.
\begin{itemize}
\item  For DBSCAN with $MinPts=3$ and $\varepsilon = 3$, we had five outliers in Figure \ref{Imagen:Figura30}(a). The smallest corresponds to day $15$. The distance is $D=0$. When $k=3$ and $\varepsilon = 3.5$, we had one outlier in Figure \ref{Imagen:Figura30}(b), day $17$. The distance is $D=2$.
\item K-Means with $k=3$ introduces the first temporal mismatch on day $15$, assigning this day to the third cluster (blue), instead of the second cluster (green), see Figure \ref{Imagen:Figura30}.
The distance is $D=0$. When $k=4$, we have a mini-cluster with days  $17$ and $18$. The distance is $D=2$. When $k=5$,  the first temporal mismatch happens for day $11$, and the distance is $D=4$. 
\item Hierchical clustering with $4$ clusters, as depicted in Figure \ref{Imagen:Figura16}, singles out day $17$. Therefore the distance is $D=2$. Figure \ref{Imagen:Figura15} considers $3$ clusters instead.  The smallest day in the smallest cluster is $15$, thus the distance is $D=0.$
\end{itemize}

We have applied these methods to $28$ different clinical datasets, quantifying for
each one the distance between the predicted and true transition days. 
We have excluded some datasets for several reasons. On three of them all the 
algorithms gave the same answer. Three more are discarded because the Hopkins
statistics is too small to support looking for clusters. The variables just fluctuate.
Three additional datasets are too complicate to analyze since they seem to present
several periods of flares and remission, and should be divided in smaller periods.
Thus, we study the remaining $19$ datasets, which present only one transition.
The results are represented in Table \ref{Table2}, where HC stands for hierarchical 
clustering (3C with $3$ clusters, 4C with $4$ clusters) and KM by K-Means. 

\begin{table}[!hbt]
\centering
\tabcolsep=0.10cm
\begin{tabular}{|c|c|c|c|c|c|c|}
\hline
\textbf{\begin{tabular}[c]{@{}c@{}}Distance to \\ transition days\end{tabular}} &
\textbf{\begin{tabular}[c]{@{}c@{}}KM \\ k=3\end{tabular}} & 
\textbf{\begin{tabular}[c]{@{}c@{}}KM\\ k=4\end{tabular}} & 
\textbf{\begin{tabular}[c]{@{}c@{}}KM \\ k=5\end{tabular}} & 
\textbf{\begin{tabular}[c]{@{}c@{}}HC \\ 3C\end{tabular}} & 
\textbf{\begin{tabular}[c]{@{}c@{}}HC \\ 4C\end{tabular}} & 
\textbf{\begin{tabular}[c]{@{}c@{}}DBSCAN \\ MinPts=3, $\varepsilon$=3\end{tabular}} \\ \hline
D15                                                               & 0                                                               & 2                                                              & 4                                                               & 0              & 2              & 0                                                                             \\ \hline
D4                                                                & 0                                                               & 1                                                              & 1                                                               & 0              & 0              & 0                                                                             \\ \hline
D9                                                                & 0                                                               & 6                                                              & 6                                                               & 9              & 9              & 2                                                                             \\ \hline
D3                                                                & 2                                                               & 2                                                              & 2                                                               & 0              & 0              & 0                                                                             \\ \hline
D1                                                                & 2                                                               & 2                                                              & 0                                                               & 2              & 2              & 0                                                                             \\ \hline
D1                                                                & 6                                                               & 10                                                             & 10                                                              & 0              & 0              & 0                                                                             \\ \hline
D16                                                               & 6                                                               & 6                                                              & 9                                                               & 19             & 19             & 15                                                                            \\ \hline
D14                                                               & 0                                                               & 0                                                              & 11                                                              & 0              & 0              & 0                                                                             \\ \hline
D5                                                                & 0                                                               & 1                                                              & 1                                                               & 0              & 0              & 0                                                                             \\ \hline
D9                                                                & 0                                                               & 5                                                              & 5                                                               & 3              & 5              & 5                                                                             \\ \hline
D19                                                               & 0                                                               & 0                                                              & 16                                                              & 17             & 0              & 17                                                                            \\ \hline
D50                                                               & 41                                                              & 41                                                             & 41                                                              & 4              & 0              & 0                                                                             \\ \hline
D1                                                                & 3                                                               & 0                                                              & 0                                                               & 0              & 0              & 0                                                                             \\ \hline
D23                                                               & 22                                                              & 22                                                             & 41                                                              & 0              & 0              & 21                                                                            \\ \hline
D17                                                               & 0                                                               & 9                                                              & 9                                                               & 0              & 0              & 12                                                                            \\ \hline
D19                                                               & 14                                                              & 14                                                             & 14                                                              & 0              & 7              & 18                                                                            \\ \hline
D11                                                               & 6                                                               & 0                                                              & 0                                                               & 0              & 0              & 4                                                                             \\ \hline
D3                                                                & 21                                                              & 14                                                             & 0                                                               & 19             & 21             & 0                                                                             \\ \hline
D12                                                               & 11                                                              & 0                                                              & 0                                                               & 11             & 0              & 10                                                                            \\ \hline
\end{tabular}
\caption{Distances to the transition day for the different algoritms. \label{Table2}}
\end{table}

\begin{table}[!hbt]
\centering
\tabcolsep=0.10cm
\begin{tabular}{|c|c|c|c|c|c|c|}
\hline

\textbf{Nº} & \textbf{\begin{tabular}[c]{@{}c@{}} KM \\ k=3\end{tabular}} &
 \textbf{\begin{tabular}[c]{@{}c@{}} KM \\ k=4 \end{tabular}} & 
 \textbf{\begin{tabular}[c]{@{}c@{}} KM \\ k=5 \end{tabular}} & 
 \textbf{\begin{tabular}[c]{@{}c@{}} HC\\ 3C \end{tabular}} & 
 \textbf{\begin{tabular}[c]{@{}c@{}} HC \\  4C \end{tabular}} & 
 \textbf{\begin{tabular}[c]{@{}c@{}} DBSCAN \\ MinPts=3, $\varepsilon$=3\end{tabular}} 
\\ \hline
\textbf{1}  & 1                                                                & 4                                                              & 6                                                               & 1              & 4              & 1                                                                             \\ \hline
\textbf{2}  & 1                                                                & 5                                                              & 5                                                               & 1              & 1              & 1                                                                             \\ \hline
\textbf{3}  & 1                                                                & 3                                                              & 3                                                               & 5              & 5              & 2                                                                             \\ \hline
\textbf{5}  & 4                                                                & 4                                                              & 4                                                               & 1              & 1              & 1                                                                             \\ \hline
\textbf{6}  & 3                                                                & 3                                                              & 1                                                               & 3              & 3              & 1                                                                             \\ \hline
\textbf{7}  & 4                                                                & 5                                                              & 5                                                               & 1              & 1              & 1                                                                             \\ \hline
\textbf{8}  & 1                                                                & 1                                                              & 3                                                               & 5              & 5              & 4                                                                             \\ \hline
\textbf{9}  & 1                                                                & 1                                                              & 6                                                               & 1              & 1              & 1                                                                             \\ \hline
\textbf{10} & 1                                                                & 5                                                              & 5                                                               & 1              & 1              & 1                                                                             \\ \hline
\textbf{11} & 1                                                                & 3                                                              & 3                                                               & 2              & 3              & 3                                                                             \\ \hline
\textbf{12} & 1                                                                & 1                                                              & 4                                                               & 5              & 1              & 5                                                                             \\ \hline
\textbf{14} & 4                                                                & 4                                                              & 4                                                               & 3              & 1              & 1                                                                             \\ \hline
\textbf{15} & 6                                                                & 1                                                              & 1                                                               & 1              & 1              & 1                                                                             \\ \hline
\textbf{18} & 4                                                                & 4                                                              & 6                                                               & 1              & 1              & 3                                                                             \\ \hline
\textbf{23} & 1                                                                & 4                                                              & 4                                                               & 1              & 1              & 6                                                                             \\ \hline
\textbf{24} & 3                                                                & 3                                                              & 3                                                               & 1              & 2              & 6                                                                             \\ \hline
\textbf{27} & 6                                                                & 1                                                              & 1                                                               & 1              & 1              & 5                                                                             \\ \hline
\textbf{28} & 5                                                                & 3                                                              & 1                                                               & 4              & 5              & 1                                                                             \\ \hline
\textbf{29} & 5                                                                & 1                                                              & 1                                                               & 5              & 1              & 4                                                                             \\ \hline
\end{tabular}
\caption{Rankings generated from the distances to the transition day in Table \ref{Table2}. \label{Table3}}
\end{table}

Based on those distances, we build the ranking presented in Table \ref{Table3}. We assign a higher position 
in the ranking to smaller distances. The smallest possible distance is $D=0$. Smallest distances
rank first. Ties are solved assigning the same position to tied algorithms and freeing the next positions
in equal number.  Following this procedure, we obtain Table \ref{Table3}.

We notice that K-Means with $k=3$, Hierarchical clustering with $3$ clusters and
DBSCAN with $\varepsilon=3$ and $MinPts=3$ perform quite well. We will use the PL method to determine the probability for each algorithm being the best, as well as the uncertainty in our choice of algorithm.
Notice that there are repetitions in the ranking, obtaining the results represented in Table \ref{Table4}.

\begin{table}[!h]
\centering
\begin{tabular}{|c|c|c|c|c|c|}
\hline
\textbf{\begin{tabular}[c]{@{}c@{}}KM\\ k=3\end{tabular}} & 
\textbf{\begin{tabular}[c]{@{}c@{}}KM\\ k=4\end{tabular}} & 
\textbf{\begin{tabular}[c]{@{}c@{}}KM\\ k=5\end{tabular}} & 
\textbf{\begin{tabular}[c]{@{}c@{}}HC\\ 3C\end{tabular}} & 
\textbf{\begin{tabular}[c]{@{}c@{}}HC\\ 4C\end{tabular}} & 
\textbf{\begin{tabular}[c]{@{}c@{}} DBSCAN \\ MinPts=3, $\varepsilon$=3\end{tabular}} \\ \hline
0,148702                                                       & 0,137208                                                       & 0,11005                                                        & 0,20365        & 0,213026       & 0,187354                                                                      \\ \hline
\end{tabular}
\label{Table4}
\caption{Results obtained applying the Plackett Luce method to the ranking in
Table \ref{Table3}.}
\end{table}

The results in Table \ref{Table4} indicate that hierarchical clustering with $4$ clusters is the algorithm
performing best, with a probability of 21.30\%.  Next, it follows K-means with $3$ clusters, with
a 20.36\% probability, which worsens increasing the number of clusters.
DBSCAN appears with probability 18.73\%.  

In spite of the narrow differences, hierarchical clustering algorithms outperform the rest due to two reasons.
First, they adapt well to small  datasets. Second, they do not require a previous knowledge of
the number of clusters, one can infer reasonable values from the tree.
DBSCAN algorithms perform worse that expected. This algorithm is devised to look for outliers, since
it does not need to place all points in a cluster, unlike K-means. However, for many parameter choices
we may find no outliers. Analyzing the results, we observe that either it gives sharp predictions or
it produces the worst predictions. This may reflect a difficulty in tuning the hyperparameters in
these algorithms,  we just used the values selected for the reference case. On the other hand,
K-Means is a classical algorithm that performs poorly in outlier detection. This is due to the
need of placing all points in a cluster, the difficulty to handle non convex clusters, and the
requirement of a priori information on the number of clusters.

Let us finally point out that the results vary with the definitions of transition days and distances
for the different algorithms. If we adopt a different definition for hierarchical clustering techniques
exploiting the cophenetic distances, then it outperforms the rest. Here we chose the simplest
definitions to illustrate the procedure.


\section{Conclusions}
\label{sec:conclusion}

Extracting information from real clinical data in an automatic fashion faces
a number of challenges, such as the unavailability of large enough amounts
of data and incompleteness of the records, for instance. For each patient
undergoing the same illness, slightly different variables may have been
monitored and time intervals between tests vary largely. Unsupervised 
clustering methods provide a tool to obtain basic information. We have 
compared the potential of DBSCAN, K-means and hierarchical clustering 
techniques to detect the presence of transitions from remission to flares 
in lupus patients using the time records of standard laboratory tests.
When faced with large numbers of data sets, one must figure out which  
clustering strategy is likely to be the best one for most of them. We show
that a  Bayesian analysis based on the Plackett-Luce model applied to 
performance rankings of clustering algorithms on a collection of model
clinical datasets may  identify the best methods with quantified uncertainty.

Ideally, one would like to go further and identify patterns representing a 
type of flare in the data, to which new datasets could be compared.
In that way, we might be able to diagnose automatically what particular 
manifestation of lupus we need to treat.
This is a challenge due to the fact that available hospital records usually 
display different variable collections measured over irregular time periods,
reacting often to emergencies.


\vskip 5mm

{\bf Acknowledgements.} This research has been partially supported by 
the FEDER /Ministerio de Ciencia, Innovación y Universidades - Agencia 
Estatal de Investigación grant No. MTM2017-84446-C2-1-R.
The authors thank Hospital Universitario Puerta de Hierro (Madrid,
Spain) for providing the clinical data.


\newpage 

{\bf Appendix: Clinical variables.}
\vskip 2mm

We list here the clinical variables involved in the test datasets under study, 
see Tables \ref{TableAp1} and \ref{TableAp3}.  The labels refer to the variables 
involved in the reference dataset used through the text to exemplify
the performance of clustering algorithms. The ranking for the Bayesian
analysis uses additional datasets containing measurements of most of these
variables. As it is to be expected from patients' records stored in hospitals,
the datasets do not always record all the variables, and the days
at which measurements are taken vary with the patient.

\begin{table}[!h] 
\centering  
\resizebox{0.9\columnwidth}{!}{%
\begin{tabular}{|c|c|c|c|}
\hline
\textbf{ \small Blood Test - Variable} & \textbf{\small Label}     &    
\textbf{ \small Urine Test - Variable} & \textbf{\small Label}   \\ \hline
\small Glucose       &  1    & \small  pH         & 33  \\ \hline
\small Urea             &  2   & \small  Density  & 34 \\ \hline
\small Creatinine    &  3    & \small  Proteins (strip) & 35  \\ \hline
\small Glomerular Filtration & 4    & \small  Glucose & 36 \\ \hline
\small Uric Acid      & 5     & \small  Ketone bodies & 37  \\ \hline
\small Cholesterol    & 6   & \small Bilirubin & 38 \\ \hline
\small Cholesterol HDL  &   7 & \small  Urobilinogen & 39 \\ \hline 
\small Cholesterol LDL   &   8  & \small  Nitrites & 40 \\ \hline
\small Triglyceride    & 9    & \small  Leukocytes & 41 \\ \hline
\small Total Proteins & 10  & \small  Red blood cell & 42  \\ \hline
\small Albumin  &  11   &  \small  Turbidity & 43     \\ \hline
\small Calcium  &  12    & \small Sediment Comment  &    \\ \hline
\small Phosphorus  & 13   & 
\small Leukocytes per field  & 44    \\ \hline
\small Sodium  & 14    &  \small G6PD quantification &   \\ \hline
\small Potassium  & 15   & 
\small Red blood cell shadow  & 45 \\ \hline
\small Chloride  & 16    & \small Hyaline cylinders &    \\ \hline
\small Bicarbonate  &  17 & \small Cell Peeling &\\ \hline
\small Iron &  18 & \small Creatinine &  46 \\ \hline
\small Total Bilirubin  &  19 & \small Proteins & \\ \hline
\small Creatine Kinase (CK) &  & \small Albumin & 47  \\ \hline
\small  Lactate dehydrogenase (LDH) & 20 
& \small  Microalbumin/Creat. Ratio & 48 \\ \hline
\small  Alanine Aminotransferase (ALTGPT)  & 21 
& \small Haptoglobin &\\ \hline
\small Aspartate Aminotransf. (ASTGOT) & 22 
& &  \\ \hline
\small Phosfatase Alkaline & 23  & & \\ \hline
\small  Gamma Glutamyltransferase (GGT) & 24 & & \\ \hline
\small Vitamin B12 &  25  & & \\ \hline
\small Folic Acid     &  26  & & \\ \hline
\small C-Reactive Protein  &  27  & & \\ \hline
\small Ferritin   & 28   & & \\ \hline
\small Transferrin & 29  & & \\ \hline
\small Transferrin Saturation & 30   & & \\ \hline
\small Complement component 3 (C3) & 31    & & \\ \hline
\small Complement component 4 (C4) & 32   & & \\ \hline
\small Thyroid Stimulating Hormone (TSH) &  & & \\ \hline
\small Parathyroid Hormone (PTHi) &  & & \\ \hline
\small v25-OH Vitamin D &  & & \\ \hline
\end{tabular}
}
\caption{Numbering of clinical variables.}
\label{TableAp1}
\end{table}

\begin{table}[!h]
\centering
\resizebox{0.7\columnwidth}{!}{%
\begin{tabular}{|c|c|}
\hline
\textbf{\small Hematology Laboratory -  Variable} & \textbf{\small Label}    
     \\ \hline
\small Leukocytes     & 49   \\ \hline
\small Neutrophils     & 50 \\ \hline
\small Lymphocytes  & 51   \\ \hline
\small Monocytes      & 52   \\ \hline
\small Eosinophils     & 53   \\ \hline
\small Basophils        & 54   \\ \hline
\small Neutrophils (Percent) & 55     \\ \hline
\small Lymphocytes  (Percent) & 56  \\ \hline
\small Monocytes   (Percent) &  57    \\ \hline
\small Eosinophils  (Percent) & 58     \\ \hline
\small Basophils  (Percent) &  59    \\ \hline
\small Immature Granulocyte (IG) count & 60   \\ \hline
\small Red blood cell & 61 \\ \hline
\small Hemoglobin & 62   \\ \hline
\small Hematocrit & 63 \\ \hline
\small Mean Corpuscular Volume (MCV) & 64   \\ \hline
\small Mean Corpuscular Hemoglobin (MCH) & 65 \\ \hline
\small Mean Corpuscular Hemoglobin Concentration (MCHC) & 66 \\ \hline \small Erythroblasts  &       \\ \hline          
\small Erythroblasts (Percent) & 67 \\ \hline
\small Red blood cell Distribution Width (RDW) & 68 \\ \hline
\small Platelets  &  69 \\ \hline
\small Platelet Distribution wWidth (PDW) & 70	 \\ \hline	
\small Reticulocytes (Percent) &  \\ \hline
\small Erythrocyte Sedimentation Rate (ESR) & 71 \\ \hline
\small Direct Coombs &  \\ \hline
\small Glucose-6-Phosphate Dehydrogenase (G6PD) quantification & \\ \hline 
\small Haptoglobin &  \\ \hline
\small Protrombrine Time & \\ \hline
\small Activated Partial Thromboplastin Time (APTT) & \\ \hline
\small Fibrinogen & \\ \hline
\small Lupus anticoagulant & \\ \hline

\textbf{\small Immunology Laboratory - Variable} & \textbf{\small Label}    
     \\ \hline
\small Anti-dsDNA antibodies & 72 \\ \hline
\small lgM Anti-cardiolipin antibodies &  \\ \hline
\small lgG Anti-cardiolipin antibodies &  \\ \hline
\small lgM AntiB2glicoprotein-I  antibodies &  \\ \hline
\small lgG AntiB2glicoprotein-I  antibodies &  \\ \hline
\end{tabular}
}
\label{TableAp3}
\caption{Numbering of clinical variables.}
\end{table}


\end{document}